\documentclass[%
 aip,
cp, 
 amsmath,amssymb,
 reprint,%
]{revtex4-2}

\usepackage{graphicx}
\usepackage{dcolumn}
\usepackage{bm}
\usepackage[mathlines]{lineno}

\usepackage[utf8]{inputenc}
\usepackage[T1]{fontenc}
\usepackage{babel}
\usepackage{mathptmx}
\usepackage{bm,multirow}

\def\b{\begin{eqnarray}}
\def\e{\end{eqnarray}}

\def\f{\frac}
\def\Bl{\Bigl}
\def\Br{\Bigr}
\def\bl{\bigl}
\def\br{\bigr}
\def\q{\quad}

\begin{document}

\title{Magnetars as Laboratories for Strong Field QED}

\author{Chul Min Kim}
\email[]{chulmin@gist.ac.kr}
\affiliation{Center for Relativistic Laser Science\char`,{} Institute for Basic Science\char`,{} Gwangju 61005\char`,{} Korea}
\affiliation{Advanced Photonics Research Institute\char`,{} Gwangju Institute of Science and Technology\char`,{} Gwangju 61005\char`,{} Korea}

\author{Sang Pyo Kim} 
\email[Corresponding author: ]{sangkim@kunsan.ac.kr}
\affiliation{Department of Physics\char`,{} Kunsan National University\char`,{} Kunsan 54150\char`,{} Korea}

\date{\today} 

\begin{abstract}
A strong electromagnetic field polarizes the vacuum and in the presence of an electric field creates pairs of a charged particle and its anti-particle.
Magnetars, highly magnetized neutron stars with magnetic field comparable to or greater than the Schwinger field, give a significant amount of the vacuum polarization and vacuum birefringence and the induced electric field can create the electron-positron pairs, which are strong field quantum electrodynamics (QED) processes. In this paper, we use a closed formula for the one-loop effective action in the presence of a supercritical magnetic field and a subcritical electric field, find the vacuum birefringence analytically and numerically, and then discuss possible measurements in magnetars.
\end{abstract}

\maketitle

\section{Introduction}

Physical phenomena in strong background fields differ from those in weak ones. The production of charged particle pairs in a strong electric field, known as the Schwinger effect, is one of the most prominent aspects of nonperturbative quantum electrodynamics (QED), and Hawking radiation from black holes is another phenomenon; both of which cannot be found by the weak field method. In quantum field theory, the pair production or the particle creation in background fields is a consequence of the vacuum persistence amplitude, the imaginary part of the effective action probed by quantum fields. In QED, the one-loop action has an imaginary part originated from the poles and gives the decay rate of the Dirac vacuum.

Heisenberg-Euler \cite{Heisenberg1936Folgerungen} and Schwinger \cite{schwinger_gauge_1951} found the one-loop effective action in a strong constant electromagnetic field by computing the interactions of the negative-energy electrons in the Dirac sea with all even numbers of photons from the background electromagnetic field, and showed that the Dirac vacuum under such a field becomes a polarized medium. When the electric field is comparable to the critical field $E_{c} = m^2c^3/e\hbar = 1.3 \times 10^{16}\, {\rm V/cm}$, electron-positron pairs are significantly produced to have the mean number of pairs as
\b
{\cal N} (E) = e^{-\frac{\pi m^2}{eE}},
\e
where we use the cgs Gaussian units with $c=\hbar =1$. Similarly, the nonlinear QED effect is significant when a magnetic field gives the Landau level spacing equal to the rest mass of electrons. The reference values for such a strong magnetic field are the critical magnetic field, also known as Schwinger field, and the corresponding critical intensity:
\b
B_c = \f{m^2c^3}{\hbar e} = 4.4 \times 10^{13}\, {\rm G}, \q I_c  = \f{B_c^2 }{8 \pi} c = 2.3 \times 10^{29}\, {\rm W/cm^2}.
\e
Note that the energy density of the critical magnetic field has the order of the rest mass of electron per unit Compton volume
\b
B_c^2 \times \f{e^2}{\hbar c} = \f{m c^2}{(\hbar/mc)^3}
\e
Physical processes in supercritical magnetic fields differ from those in weak fields or without fields, such as the curvature synchrotron radiation and QED cascade of pairs \cite{Meszaros1992High,Battesti:2012hf,Harding2006Physics}, not to mention the electron-positron pair production in astrophysics \cite{ruffini2010electron}.

In laboratory, the CPA invented by Mourou and Strickland \cite{Strickland1985Compression} boosted the intensity of ultra-intense lasers, and the extreme intensity of $ I=1.1 \times 10^{23}\, {\rm W/cm^2}$ was achieved recently by CoReLS \cite{Yoon2021Realization}. As plasma mirrors may enhance the laser intensity without limit in principle \cite{Thaury2007Plasma}, and the critical field intensity will be achieved in the future. Ultra-intense lasers are opening a window for laboratory astrophysics (\cite{Kim2019Astrophysics} and references therein). Recently, the ATLAS experiments using accelerators found an evidence of the light-by-light scattering from heavy ion collisions, in which the field strength was estimated to reach $10^{23}\,{\rm  V/cm}$ \cite{ATLASCollaboration2017Evidence}.

In astrophysics, neutron stars can have extremely high magnetic fields, whose theoretical upper bound Chandrasekhar and Fermi found by applying the virial theorem of magnetohydrostatic equilibrium \cite{chandrasekhar1953problems} (for review and references, see \cite{Shapiro1983Black})
\b
\f{4 \pi R_{\rm N}^3}{3} \times \f{B^2}{8 \pi} \leq G \f{M_{\rm N}^2}{R_{\rm N}} \Rightarrow B \leq \Bl(\f{M_{\rm N}}{1.4 M_{\odot}} \Br) \Bl(
\f{10\, {\rm km}}{ R_{\rm N}} \Br)^2,
\e
where $G$, $R_{\mathrm N}$, and $M_{\mathrm N}$ are the gavitational constant, the radius of the neutron star, and its mass, respectively. Though the magnetic flux conservation during the stellar collapse associates $10^{12}\, {\rm G}$ with neutron stars \cite{woltjer1964x}, the $\alpha-$ and $\omega-$dynamos in proto-neutron stars can generate magnetic fields of order $10^{15}\, {\rm G}$ or stronger \cite{thompson1993neutron}. Such highly magnetized neutron stars, known as magnetars, have been observed \cite{vasisht1997discovery}. Table \ref{Tab1} lists some magnetars with supercritical magnetic fields from McGill Catalog \cite{olausen_mcgill_2014}, and the physical, astrophysical properties of magnetars and their observations are reviewed in Refs.~\cite{kaspi2017magnetars,enoto_observational_2019}. Recently space missions have been proposed to probe the QED regime of highly magnetized neutron stars: Compton Telescope \cite{wadiasingh2019magnetars} and eXTP \cite{santangelo2019physics}.

\begin{table}
\caption{\label{Tab1} Compact stars with supercritical magnetic fields from McGill magnetar catalog \cite{olausen_mcgill_2014}}
\begin{ruledtabular}
 \begin{tabular}{llll}
 {\bf Name} & {\bf Period P} ${\rm (s)}$ & {\bf Period Change $\dot{P}$} $10^{-11} {\rm (s s^{-1})}$ & ${\bf B/B_C}$ \\
 \hline
SGR 0526-66 & 8.0544(2) & 3.8(1) & 1.35 \\
IRXS J170849.0-400910  & 11.003027(1) & 1.91(4) & 1.11  \\
CXOU J171405.7-381031 0526-66 & 3.825352(4) & 6.40(5) & 1.21 \\
SGR 1806-20 & 7.547728(17) & 49.5 & 4.83  \\
IE 1841-045 & 11.782898(1)& 3.93(1) & 1.67 \\
SGR 1900+14 & 5.19987(7) & 9.2(4) & 1.69
 \end{tabular}
\end{ruledtabular}
\end{table}

\section{QED Action in Supercritical Magnetic Field and Subcritical Electric Field}

It is commonly believed that the Heisenberg-Euler and Schwinger QED action substituted by a local and temporal configuration of electromagnetic field (locally constant field (LCF)) can be applied to a spatially and temporally varying electromagnetic field when the characteristic length is larger than the Compton length, and the characteristic time scale is longer than the Compton time. However, the magnetic length $\lambda_B = \lambda_C (B_c/B)^{1/2}$ is shorter in supercritical magnetic fields than the Compton length $\lambda_C$. In Ref. \cite{kim2006schwinger}, the LCF approximation for a Sauter field gives a good approximation for the electron-positron pair production when $\lambda_C/L \ll (\lambda_C/\lambda_E)^4$ with
the electric field length $\lambda_E = \lambda_C (E_c/E)^{1/2}$, while it is no longer a good approximation when $ (\lambda_C/\lambda_E)^4 \ll \lambda_C / L \ll (\lambda_C/\lambda_E)^3$. The QED phenomena in strong fields, such as in heavy atoms and ultra-intense laser pulses, should properly take into account of the field length as well as the characteristic length scale \cite{greiner2012quantum}.

In the case of the Sauter field, a localized electric or magnetic field, the LCF approximation for the QED action is a good approximation when the length scale is much larger than the Compton length and the electric or magnetic length \cite{kim_effective_2008,kim_effective_2010,kim2011qed}. The magnetars in Table \ref{Tab1} have macroscopic length and time scales, which are extremely larger than the Compton length (and time) and the magnetic length. So it is legitimate to use the Heisenberg-Euler and Schwinger QED action.

The QED action at one-loop in a supercritical magnetic field and a subcritical electric field, $\beta =eB/m^2  \gg 1$ and $\epsilon = eE/m^2  \leq 1$, was obtained by Ritus \cite{ritus_lagrangian_1976} and Dittrich \cite{dittrich_evaluation_1979} as
\b \label{one-loop}
{\cal L}^{(1)}= \f{m^4}{24 \pi^2} \beta^2 \Bl[ \ln \Bl(\f{ \beta }{\gamma \pi} \Br) + \f{6}{\pi^2} \zeta'(2)  + \cdots \Br],
\e
where and hereafter $\alpha$, $\gamma$, and $\zeta$ are the fine structure constant, the Euler's constant, and the Riemann zeta function, respectively. In Ref.~\cite{kim_quantum_2019}, the one-loop action (\ref{one-loop}) is also obtained by using the gamma-function regularization in the in-out formalism.
The QED action at two-loop was calculated by Ritus \cite{ritus_lagrangian_1976} as
\b
{\cal L}^{(2)} = \frac{3\alpha}{4\pi} \f{m^4}{24 \pi^2} \beta^2 \Bl[ \ln \Bl(\f{ \beta}{\gamma \pi} \Br) + a_2 \Br],
\e
where $a_2$ is a real constant. Thus, the QED action up to the two-loop level is
\b
{\cal L}^{(1)}+{\cal L}^{(2)} = \f{m^4}{24 \pi^2}\beta^2 \Bl[ \Bl(1+\f{3 \alpha}{4 \pi}\Br)  \ln \Bl(\f{ \beta}{\gamma \pi} \Br) +
\f{6}{\pi^2} \zeta'(2)+ \frac{3\alpha}{4\pi} a_2 + \cdots \Br].
\e
The two-loop action are  smaller by a factor of $\alpha$  than the one-loop action, and thus the one-loop QED action is still a good approximation in the supercritical magnetic fields.

In general, the one-loop QED action for a constant electromagnetic field is given by
\b \label{HES}
{\mathcal L^{(1)}} (a,b) = - \f{1}{8 \pi^2} \int_{0}^{\infty} ds \f{e^{-m^2 s}}{s^3}
\Bl[(esa) (ebs) \coth(eas) \cot(ebs) - 1 - \f{(es)^2}{3} (a^2 - b^2) \Br],
\e
where $a$ and $b$ are gauge and Lorentz invariant quantities:
\b
a = \sqrt{\sqrt{{\cal F}^2 + {\cal G}^2} + {\cal F}}, \quad b = \sqrt{\sqrt{{\cal F}^2 + {\cal G}^2} - {\cal F}}.
\e
Here, ${\cal F} = F_{\mu \nu} F^{\mu \nu}/4=(\vec{B}^2-\vec{E}^2)/2$ is the Maxwell scalar, and ${\cal G} = F_{\mu \nu} F^{* \mu \nu}/4=-\vec{E}\cdot\vec{B}$ is the psuedo-scalar; from now on, we switch to the Lorentz-Heaviside units with $c=\hbar=1$.
Noting that ${\cal L}^{(1)} (\pm a, \pm b) = {\cal L}^{(1)} ( a, b)$, the effective action is in powers of $a^2$ and $b^2$.
Also, the QED action has the electromagnetic duality $\vec{B} \leftrightarrow i \vec{E}$ \cite{kim2011qed,kim_quantum_2019} since ${\cal L}^{(1)} (a, b) = {\cal L}^{(1)} (ib, -ia)$. One cannot express the spectral function in a single $\coth$ since $\coth(eas) \cot(ebs) \neq \coth (e (a+ib)s)$.

When the pseudo-scalar is null, i.e., ${\cal G} = 0$ ($b = 0, a = \sqrt{2 {\cal F}}$), the one-loop action reads
\b
{\cal L}^{(1)}(a,0)= -\f{1}{8\pi^2} \int_{0}^{\infty} ds \f{ e^{-m^2 s}}{s^3}
\Bl[(esa) \coth(eas) - 1 - \f{(esa)^2}{3} \Br].
\e
The closed form of this one-loop action was obtained by Dittrich \cite{dittrich_one-loop_1976} and shown exact for an arbitrary value of $\bar{a}={m^2}/{2ea}$ by Kim and Lee \cite{kim_quantum_2019} using the in-out formalism:
\begin{equation}
	\label{act supB}
	\mathcal{L}^{(1)}(a,0)=\frac{m^{4}}{8\pi^{2}\bar{a}^{2}}\left[\zeta'(-1,\bar{a})-\frac{1}{12}+\frac{\bar{a}^{2}}{4}-\left(\frac{1}{12}-\frac{\bar{a}}{2}+\frac{\bar{a}^{2}}{2}\right)\ln\bar{a}\right],
\end{equation}
where $\zeta'(-1,\bar{a})=d\zeta(s,\bar{a})/ds|_{s=-1}$, and $\zeta(s,\bar{a})$
is the Hurwitz zeta function.

In the context of strongly magnetized neutron stars, we consider a supercritical magnetic field and a subcritical electric field, in which the condition of $a \gg b$ holds. Then, we can expand the effective action (\ref{HES}) as a power series in $b^2$:
\b \label{act exp}
{\mathcal L}^{(1)} (a, b) = \sum_{n= 0}^{\infty} \f{\partial^{2n}}{\partial b^{2n}} {\mathcal L}^{(1)} (a, b) \Bl|_{b = 0} \f{b^{2n}}{(2n)!}.
\e
Note that this series is gauge- and Lorentz-invariant. Similarly, Heyl and Hernquist expanded the QED action with respect to ${\cal G} $ \cite{heyl_analytic_1997,heyl_birefringence_1997}.

\section{Vacuum birefringence in a magnetic field}

\begin{figure}
	\centering
	\includegraphics[width=0.25\textwidth]{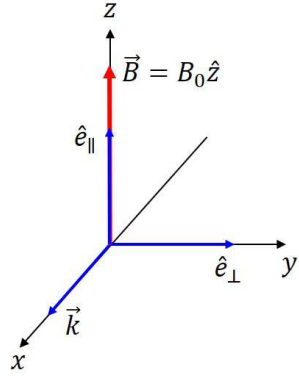}
	\caption{Configuration of the background magnetic field and the probe field.
		The background field is along the $z$-axis, and the vector $\vec{k}$ along the $x$-axis
		denotes the propagation vector of the probe field. The unit vectors
		$\hat{e}_{\parallel}$ and $\hat{e}_{\perp}$ are the polarization vectors associated with the refractive indices $n_{\parallel}$ and $n_{\perp}$, respectively.}
	\label{fig:field_conf90}
\end{figure}

Strong magnetic fields, albeit incapable of producing electron-positron pairs, can turn the vacuum into an optical medium, and thus an electromagnetic wave  propagating through a magnetized vacuum, called the probe field below, can undergo birefringence. It is a representative phenomenon of nonlinear QED, revealing the striking difference between the classical and quantum vacua. Recently, the measurement of vacuum birefringence by using an ultra-intense laser and an x-ray free electron laser has become a top research topic for upcoming laser facilities \cite{Schlenvoigt2016Detecting,Shen2018Exploring}, albeit the available field strength is much smaller than the critical strength. To study the vacuum birefrigence around and above the critical strength, one should resort to the compact astrophysical objects such as neutron stars, magnetars, white dwarfs, and black holes, which can produce extremely strong magnetic and electric fields. A polarimetric observation of the x-rays from such astrophysical bodies would provide the information to prove and understand the vacuum birefringence in supercritical fields \cite{mignani_evidence_2017,heyl_strongly_2018,caiazzoVacuumBirefringenceXray2018, caiazzo2021highly}. In this regard, we derive the refractive indices for a probe field by using the one-loop effective action presented in the previous section. For simplicity, we restrict our derivation to the case that only magnetic field is present, and the probe field propagates across the magnetic field lines, as shown in Fig.~\ref{fig:field_conf90}. An account of the more general cases are being prepared.

In a background electromagnetic field, the vacuum acquires the effective action as
\b
{\cal L}_{\rm eff} &=& - \frac{1}{2} \bl(b^2 - a^2 \br) + {\cal L}^{(1)} (a, b).
\e
The polarization and magnetization can be obtained by considering the variation of the effective action with respect to the electric and magnetic fields, and, thus, the permittivity and permeability tensors ($\epsilon_{ij}$ and $ \mu_{ij}$) can be obtained \cite{Heisenberg1936Folgerungen,berestetskii_quantum_1982}:
\b
\vec{D} = \f{\partial{\cal L}_{\rm eff}}{\partial \vec{E}} = \vec{E} + \vec{P}, \quad \vec{P} = \f{\partial{\cal L}^{(1)}}{\partial \vec{E}}, \quad
D_i = \epsilon_{ij} E_j \nonumber\\
\vec{H} = - \f{\partial{\cal L}_{\rm eff}}{\partial \vec{B}} = \vec{B} - \vec{M}, \quad \vec{M} = \f{\partial{\cal L}^{(1)}}{\partial \vec{B}}, \quad
H_i = \mu^{-1}_{ij} B_j.
\label{eq:linear_response}
\e
Strictly speaking, the permittivity and permeability tensors obtained in this way are valid for DC fields only, but they can be good approximation for the probe field of which photon energy is much smaller than the electron's rest energy \cite{berestetskii_quantum_1982}: keV hard x-ray probe fields can be treated in this way. By decomposing the electric and magnetic fields into the DC component from the background field ($\vec{B}_0$, $\vec{E}_0$) and the fluctuating one from the probe field ($\vec{B}_p$, $\vec{E}_p$),
\begin{equation}
	\vec{B} = \vec{B}_0 + \vec{B}_p , \quad \vec{E} = \vec{E}_0 + \vec{E}_p ,
\end{equation}
one can correspondingly decompose the permittivity and permeability tensors \cite{adler_photon_1971}. Below, we consider only the tensors relevant to the probe field.

For the field configuration in Fig.~\ref{fig:field_conf90}, the permittivity and permeability tensors are calculated as

\begin{equation}
	\epsilon_{ij}=\delta_{ij}(1-\hat{P}\mathcal{L}^{(1)}(a,b))+\delta_{i3}\delta_{j3}B_{0}^{2}\hat{A}^{2}\mathcal{L}^{(1)}(a,b),
\end{equation}

\begin{equation}
	\mu_{ij}^{-1}=\delta_{ij}(1-\hat{P}\mathcal{L}^{(1)}(a,b))-\delta_{i3}\delta_{j3}B_{0}^{2}\hat{P}^{2}\mathcal{L}^{(1)}(a,b),
\end{equation}

where $\hat{P}$ and $\hat{A}$ are differential operators defined as

\begin{equation}
	\hat{P}=\frac{a\cdot\partial_{a}-b\cdot\partial_{b}}{a^{2}+b^{2}} \bigg\rvert_{\vec{B}_0 , \vec{E}_0}=\frac{\partial}{\partial\mathcal{F}}\bigg\rvert_{\vec{B}_0 , \vec{E}_0},\quad\hat{A}=\mathrm{sgn}(\mathcal{G})\frac{b\cdot\partial_{a}+a\cdot\partial_{b}}{a^{2}+b^{2}}\bigg\rvert_{\vec{B}_0 , \vec{E}_0}=\frac{\partial}{\partial\mathcal{G}}\bigg\rvert_{\vec{B}_0 , \vec{E}_0}.
\end{equation}

For a pure magnetic field as in Fig.~\ref{fig:field_conf90}, $\mathcal{L}^{(1)} (a,0)$ might be considered sufficient to calculate $\epsilon_{ij}$ and $\mu^{-1}_{ij}$ because $b=0$ in such a case. However, the probe field does not satisfy the condition $b=0$, and, therefore, the expansion (\ref{act exp}) should be used for the calculation. Once $\epsilon_{ij}$ and $\mu_{ij}^{-1}$ are available, the characteristic refractive indices and the associated polarization vectors can be obtained according to the standard procedure in optics \cite{Born2019}.

In the weak field limit ($B/B_c,  E/E_c \ll 1$), the perturbative expansion of (\ref{HES}) in powers of $eas$ and $ebs$ yields the lowest-order action as
\b
\label{Leff_wf}
{\cal L}_{\rm eff}&=&  - \frac{1}{2} \bl(b^2 - a^2 \br) + \f{ 2 \alpha^2}{45 m^4} \Bl[ \bl(b^2 - a^2 \br)^2+ 7\bl(ab \br)^2  \Br] + \cdots \nonumber\\
&=& - \frac{1}{2} \bl(\vec{B}^2 - \vec{E}^2 \br) +  \f{ 2\alpha^2}{45 m^4} \Bl[ \bl(\vec{B}^2 - \vec{E}^2 \br)^2 + 7\bl(\vec{E} \cdot \vec{B} \br)^2  \Br] + \cdots.
\e
Then, the dielectric permittivity and magnetic permeability tensors are obtained as \cite{klein_birefringence_1964,adler_photon_1971}
\b
\epsilon_{ij} &=& \delta_{ij} \left(1 - \f{ 8 \alpha^2}{45 m^4}\vec{B}^2   \right) + \f{ 28 \alpha^2}{45 m^4} B_i B_j, \nonumber\\
\mu^{-1}_{ij} &=& \delta_{ij} \left(1 - \f{ 8 \alpha^2}{45 m^4}\vec{B}^2   \right) - \f{ 16 \alpha^2}{45 m^4} B_i B_j,
\e
from which the refractive indices exhibiting the vacuum birefringence in a magnetic field is obtained as
\b
n_{\parallel} = 1 + \frac{14}{45} \f{\alpha^2 B^2 }{ m^4}= 1 + \frac{7\alpha}{90\pi} \left( \frac{\vec{B}}{B_c} \right)^2, \quad
n_{\perp} = 1 + \frac{8}{45} \f{\alpha^2 B^2 }{ m^4}= 1 + \frac{2\alpha}{45\pi} \left( \frac{\vec{B}}{B_c} \right)^2.
\e
The polarization vector associated with each refractive index is shown in Fig.~\ref{fig:field_conf90}.

In an arbitrarily strong magnetic field and a weak electric field ($B/B_c \ge 1,  E/E_c \ll 1$), we use the one-loop QED actions (\ref{act exp}) and (\ref{act supB}). The resulting refractive indices are given as
\begin{equation}
	n_{\parallel}=\sqrt{\frac{1-X+Y}{1-X}},\quad n_{\perp}=\sqrt{\frac{1-X}{1-X-Z}},
\end{equation}
where

\begin{equation}
	X=-\frac{\alpha}{6\pi}\left[1-6\ln(2\pi)\bar{a}+6\bar{a}^{2}+2\ln\bar{a}-6\bar{a}\ln\bar{a}+12\bar{a}\ln\Gamma(\bar{a})-24\zeta'(-1,\bar{a})\right],
\end{equation}

\begin{equation}
	Y=-\frac{\alpha}{6\pi\bar{a}}\left[1+\bar{a}-6\ln(2\pi)\bar{a}^{2}+6\bar{a}^{3}-6\bar{a}^{2}\ln\bar{a}+12\bar{a}^{2}\ln\Gamma(\bar{a})+2\bar{a}\psi(\bar{a})-24\bar{a}\zeta'(-1,\bar{a})\right],
\end{equation}

\begin{equation}
	Z=\frac{\alpha}{3\pi}\left[1+3\left(1+\ln(2\pi)\right)\bar{a}-6\bar{a}^{2}-3\bar{a}\ln\bar{a}-6\bar{a}\ln\Gamma(\bar{a})+6\bar{a}^{2}\psi(\bar{a})\right].
\end{equation}
The symbols  $\Gamma(\bar{a})$ and $\psi(\bar{a})$ are the gamma and digamma functions, respectively. The polarization vectors remain the same. Similar results were obtained by Heyl \cite{heyl_birefringence_1997,heyl_analytic_1997}, but our scheme can be extended straightfowardly to the cases with $b \neq 0$ and $E/E_c \ll 1$. A full account of the scheme is under preparation.

\begin{figure}
	\centering
	\includegraphics[width=1\textwidth]{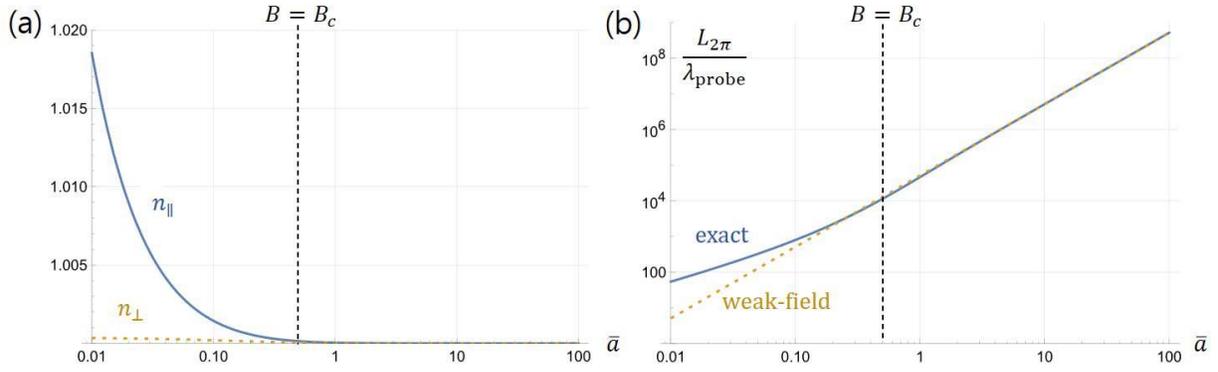}\caption{Vacuum birefringence in a magnetic field: (a) refractive indices $n_{\parallel}$ and $n_{\perp}$ as a function of the parameter $\bar{a}=(B_{c}/B)/2$; (b) the propagation length to obtain a phase difference of $2\pi$ as a function of $\bar{a}$, in units of the probe field's wavelength. The length was calculated by using the exact action (\ref{act supB}) with the expansion (\ref{act exp}) and the weak-field approximation (\ref{Leff_wf}). The vertical dashed lines marks the position corresponding to $B=B_c$.}
	\label{fig:n}
\end{figure}

Figure \ref{fig:n} exemplifies the vacuum birefringence in a magnetic field for the range from $B = 50 B_c$ ($\bar{a}=0.01$) to $B = 0.005 B_c$ ($\bar{a}=100$). In Fig.~\ref{fig:n}(a), the refractive indices begins to differ from the vacuum value 1 significantly around $B\sim B_c$, and the refractive index associated with the parallel polarization shows the dominant change, while the other index varies marginally. This behavior is the opposite to that of the birefringence due to real charged particles. In Fig.~\ref{fig:n}(b), the propagation length necessary to build a phase difference of $2\pi$, denoted by $L_{2\pi}$ is shown in units of the probe field's wavelength: $L_{2\pi}/{\lambda_{\mathrm{probe}}} = 1/{|n_{\parallel}-n_{\perp}|}$. Again, a significant departure from the weak field result is observed for $B \gtrsim B_c$.

For a comparison, we consider the laboratory demonstration using 1-keV probe photons from a x-ray free electron laser and a background field with an intensity of $I=2.3\times 10^{23} \, \mathrm{W/cm^2}$ ($B=10^{-3} B_c$) from an ultra-intense laser. Then, the probe wavelength is $1.2\,\mathrm{nm}$, and $L_{2\pi}$ is $16\,\mathrm{m}$. Considering that the laser's focal region has a longitudinal length of $\mathrm{\mu m}$ order, we find that the probe photons should traverse the focal region over million times. However, when $B\sim B_c$ as near magnetars and neutron stars, $L_{2\pi}$ can be as short as $10\, \mathrm{\mu m}$, extremely shorter than the macroscopic scale relevant with those astrophysical objects. Because such objects can have a wide variation in the distribution the magnetic field strength, the polarization states of the probe field will be accordingly distributed \cite{heyl_strongly_2018, caiazzoPolarizationAccretingXray2021, caiazzoPolarizationAccretingXray2021a}.

\section{Conclusion}

In QED, the effective action in a strong electromagnetic field predicts the vacuum polarization and the electron-positron pair production in a strong electric field: Heisenberg-Euler and Schwinger action in a constant electromagnetic field is the most well-known, exact one-loop effective action.  The effects of the vacuum polarization and pair production in various configurations of electromagnetic fields have been studied theoretically, and laboratory experiments have been proposed to observe these nonlinear QED effects. At present, the highest strength of the electric and magnetic fields from ultra-intense lasers is lower by three orders than the critical field. Thus, electron-positron pair production is exponentially small, and possible measurement of vacuum birefringence requires ingenious, state-of-the-art technology.

On the other hand, the universe provides us with sources of strong magnetic fields such as highly magnetized neutron stars and magnetars in particular, which have field strength even larger than the critical (Schwinger) field. In these ultra-strong magnetic fields, the vacuum polarization effects such as vacuum birefringence, light-light scattering, and photon splitting can be measured directly or indirectly. The characteristic length and time scales are extremely larger than the Compton length, time and the magnetic field length, to which the Heisenberg-Euler and Schwinger QED action can be applied as a good approximation.

In this paper, we have used a new analytic expression for the Heisenberg-Euler and Schwinger QED action to obtain the vacuum birefringence in supercritical magnetic fields. The expression of QED action expanded in powers of the Maxwell pseudoscalar is calculated in a regime of supercritical magnetic fields. The new expression in terms of the Hurwitz-zeta function was found by Dittrich in a pure magnetic field or an electric field in the asymptotic region~\cite{dittrich_one-loop_1976}, which was later confirmed to be exact for all strengths of magnetic fields in the in-out formalism~\cite{kim_quantum_2019}. According to the new expression, the refractive index for the parallel polarization significantly increases in supercritical magnetic fields, while that for the perpendicular polarization increases by a small amount, as shown in Fig.~\ref{fig:n}. As a consequence, if the length needed for a phase difference of $2\pi$ is calculated with the new expression for a supercritical magnetic field, it is significantly longer than is calculated with the weak-field expression. The theoretical prediction in this paper can be employed to understand the field strength and the direction of magnetic fields in magnetars or highly magnetized neutron stars.

\begin{acknowledgments}
The authors were benefited from helpful discussions with Dong Hoon Kim on highly magnetized neutron stars and with Hyun Kyu Lee on stability of compact stars and the Maxwell theory in curved spacetime. They also would like to thank Pisin Chen, Sung-Won Kim, Bum Hoon Lee and Remo Ruffini for useful discussions. This work was supported in part by the Institute for Basic Science (IBS) under IBS-R012-D1. The work of SPK was also supported in part by National Research Foundation of Korea (NRF) funded by the Ministry of Education (2019R1I1A3A01063183).
\end{acknowledgments}

\bibliography{ref_VBir}

\end{document}